# Elastic Properties of Liquid Surfaces Coated with Colloidal Particles

Edward Bormashenko[*,a,b], Gene Whyman[a], Oleg Gendelman[c]

[a]*Ariel University, Physics Department, P.O.B. 3, 40700, Ariel, Israel*

[b]*Ariel University, Chemical Engineering and Biotechnology Department, P.O.B. 3, 40700, Ariel, Israel*

[c]*Faculty of Mechanical Engineering, Technion – Israel Institute of Technology, Haifa 32000, Israel.*

[*]Corresponding author:

Edward Bormashenko

Ariel University, Physics Department, Chemical Engineering and Biotechnology Department

P.O.B. 3, Ariel 40700, Israel

Phone: +972-3-906-6134

Fax: +972-3-906-6621

E-mail: edward@ariel.ac.il

**Abstract**

The physical mechanism of elasticity of liquid surfaces coated with colloidal particles is proposed. It is suggested that particles are separated by water clearings and the capillary interaction between them is negligible. The case is treated when the colloidal layer is deformed normally to its surface. The elasticity arises as an interfacial effect. The effective Young modulus of a surface depends on the interfacial tension, equilibrium contact angle, radius of colloidal particles and their surface density. For the nanometrically scaled particles the line tension becomes essential and has an influence on the effective Young modulus.

**1. Introduction**

Colloidal particles attached to liquid surfaces are abundant in many products and processes, including crude-oil emulsions, food foams and flotation, and were studied intensively in the last decade [1,2]. When colloidal particles are attached to liquid/liquid interfaces they can work as surfactants, promoting stabilization of emulsions [3,4]. Colloidal particles attached to liquid surface gave rise to colloidosomes, i.e. selectively permeable capsules demonstrating a potential for drug delivery [5]. When attached to liquid droplets' surfaces, colloidal particles allowed manufacturing the so-called liquid marbles, which are non-stick droplets, presenting an alternative to the lotus-inspired superhydrophobicity [6-13]. Liquid marbles demonstrated a diversity of promising applications, including encapsulation, microfluidics, cultivation of microorganisms, gas-sensing, miniaturized synthesis and water storing [14-20].

Solid particles attached to liquid surface initiate a broad diversity of physical effects including capillarity-induced self-assembly and particle-assisted wetting [21-23]. One of the most interesting phenomena related to the ensembles of colloidal particles located at the liquid/air interfaces is the quasi-elastic behavior of such systems [24-26]. Vella *et al.* proposed the physical mechanism of elasticity demonstrated by the layer of colloidal particles covering a liquid surface. In our paper we introduce an alternative mechanism of elasticity of these layers.

**2. Results and discussion**

It was demonstrated recently that liquid surfaces coated with colloidal particles behave as two dimensional elastic solids (and not liquids) when deformed [24-26]. For example, liquid marbles do not coalesce when pressed one to another, as shown in Figure 1. When liquid marbles collide they do not coalesce, but demonstrate quasi-elastic collision depicted in Figure 2. It was also demonstrated that the layers built of monodisperse colloidal particles can support anisotropic stresses and strains [26]. The observed pseudo-elastic properties of liquid surfaces coated with colloidal particles call for explanation.

Vella *et al.* treated the collective behavior of a close packed monolayer of non-Brownian colloidal particles placed at a fluid-liquid interface [26]. In this simplest case, however, the close-packed monolayers may be characterized using an effective Young's modulus and Poisson ratio [26]. These authors proposed an expression for the effective Young modulus $E$ of the "interfacial particle raft" in the form[26]

$$E \cong \frac{1-\nu}{1+\phi}\frac{\gamma}{d} \qquad (1)$$

where $\gamma$ is the surface tension of liquid, $d$ is the diameter of solid particles, $\nu$ is the Poisson ratio of solid particles, and $\phi$ is the solid fraction of the interface. They concluded that the elastic properties of such an interface are not dependent on the details of capillary interaction between particles [26].

The model presented by Vella *et al*. implies close packing of elastic spheres, coating a liquid marbles. However, in the case of liquid marbles, e.g., the surface is not completely coated by solid particles, as shown in Figs. 3A-B, representing ESEM images of liquid marbles coated with lycopodium (for the detailed discussion of wetting of lycopodium particles see Ref. 27) and with polyvinylidene fluoride (PVDF) particles. Water clearings separating particles and aggregates of particles are distinctly seen [28]. It should also be emphasized that liquid marbles demonstrate "transversal" (normal to their surface) and not lateral pseudo-elasticity when being pressed or collide, whereas the model developed in Ref. 26 treats only the lateral elasticity of surfaces, coated with solid particles.

We propose an alternative mechanism of pseudo-elasticity of liquid surface covered by solid particles, which are not close-packed, explaining the "transversal" elasticity of surfaces, depicted in Fig. 4. As it will be seen below, the elasticity in this case is caused by the change in the liquid area under deformation, as it also occurs in Ref. 26.

Consider two media (one of which may be vapor) in contact (Fig. 5). The plane boundary between them is characterized by the specific surface energy (or the surface tension) $\gamma_{1,2}$. At this surface floats a spherical body of the radius *R* modeling a colloidal particle. Note that this flotation is due to surface forces but not to gravity since the size of a colloidal particle is much lower than the capillarity length $\sqrt{\gamma_{1,2}/g|\rho_1 - \rho_2|}$ ($\rho_1$ and $\rho_2$ are the corresponding densities). Besides $\gamma_{1,2}$, these forces are characterized by the surface tensions $\gamma_1$ and $\gamma_2$ at the corresponding solid-fluid interfaces (see Fig. 5).

The total surface energy *U* is given by:

$$U = 2\pi R(R - h)\gamma_1 + 2\pi R(R + h)\gamma_2 - \pi\gamma_{1,2}(R^2 - h^2) + 2\pi\Gamma\sqrt{R^2 - h^2} \quad (2)$$

The first and second terms represent the surface energies of the solid-fluid interfaces, while the third one describes the energy of the "disappearing" area due to the solid body. Also the so-called line tension, $\Gamma$, of the triple line, neighboring two fluid and one solid media, is included in Eq. (2) since it may be important for very small particles [29]. As is known, the flotation of strongly hydrophilic nanoparticles may be explained by considering the line tension only.

The equilibrium depth of immersion $h_0$ can be found by differentiation of (2)

$$\frac{\gamma_2 - \gamma_1}{\gamma_{1,2}} + \frac{h_0}{R} - \frac{\Gamma}{\gamma_{1,2}R} \cdot \frac{h_0}{(R^2 - h_0^2)^{\frac{1}{2}}} = 0 \quad (3)$$

In the widespread case when $\Gamma \ll \gamma_{1,2}R$, one can write down the approximate explicit solution of (3), like it was done in Ref. 29. To the first order in the dimensionless parameter

$$\kappa = \frac{\Gamma}{\gamma_{1,2}R} \quad (4)$$

the result for equilibrium depth of floating is:

$$h_0 = R(\cos\theta + \kappa\cot\theta) \quad (5)$$

Here we also use notation for the Young contact angle

$$\cos\theta = \frac{\gamma_1 - \gamma_2}{\gamma_{1,2}} \quad (6)$$

To provide a correct asymptotic behavior of expression (5), one should require $\kappa/\sin\theta \ll 1$. In order to study the oscillatory response of the surface covered by the colloidal particles, we obtain for the force component acting on the solid body in the case of small deviations from the equilibrium depth $h_0$

$$F_h = -\frac{dU}{dh} = -2\pi\gamma_{1,2}(1 - \frac{\kappa}{\sin^3\theta}) \cdot (h - h_0) \quad (7)$$

which is of a Hooke law form.

As it follows from (7), our model system is a harmonic oscillator with the eigen-frequency

$$\omega = \sqrt{\frac{2\pi\gamma_{1,2}(1 - \frac{\kappa}{\sin^3\theta})}{m}} \quad (8)$$

($m$ is the mass of the floating body; Exp.(8) is obviously applicable when $\kappa/\sin^3\theta < 1$). Note that elastic properties don't depend on the surface tensions $\gamma_1$ and $\gamma_2$ which determine only the equilibrium of the system in (5), (6). Exp. (8) may be rewritten in the form revealing an explicit dependence of the eigen-frequency on the radius of a particle $R$:

$$\omega = l_{ca}\sqrt{\frac{3}{2}\frac{g}{R^3}(1 - \frac{\kappa}{\sin^3\theta})} \quad (8a)$$

where $l_{ca} = \sqrt{\gamma_{1,2}/\rho_p g}$, $\rho_p$ is the density of a colloidal particle.

Now let $n$ be the surface density of colloidal particles on the area $S$ of the boundary (the surface is supposed to be flat). Under the deviation $h - h_0$, the total elastic force on the area $S$ is $nSF_h$. According to the Hooke law

$$\frac{nSF_h}{S} = E\frac{h - h_0}{|h_0|}$$

and for the effective Young modulus $E$ it follows on account of (7)

$$E = 2\pi n |h_0| \gamma_{1,2} \left(1 - \frac{\kappa}{\sin^3\theta}\right) \approx 2\pi n R \gamma_{1,2} |\cos\theta - \kappa \cot^3\theta| \qquad (9)$$

Note, that here the deformation is connected with the interface change but not with the deformation of particles. The whole approach is applicable for dilute colloids or colloids with a weak interparticle interaction. This assumption is justified for colloidal particles with the characteristic dimensions of particles $R \leq 5\,\mu$m, as shown in Ref. 21. Such particles do not deform the water/vapor interface; this leads to a negligible capillary interaction [21]. Exp. 9 will predict the effective elastic modulus for both of flat and curved surfaces coated by colloidal particles, in the case when the characteristic dimension of the deformed area $L$ is much larger than that of a particle $R$ (see Fig. 4). It should be stressed that Eq. 9 supplies the upper limit of the effective elastic modulus, because it assumes the simultaneous contact of a plate with all of colloidal particles.

An example of numerical estimation of (9) may be given using Fig. 3A. The radius of particles is of order $R \sim 1\,\mu$m; thus, taking into account the most reasonable estimation of line tension $\Gamma \sim 10^{-10}$N [30] and water surface tension $\gamma_{1,2} \sim 10^{-1}$ N/m, we see that according to (4) $\kappa \sim 10^{-3}$, and line tension is negligible. From Fig. 3A, $n \sim 10^9$m$^{-2}$, for lycopodium $\theta \sim 120°$, and from (9) a realistic estimation follows $E \sim 100$ Pa that is two orders of magnitude lower than that following from (1) in the case of close packed microparticles. Recall that our model implies a "dilute" distribution of particles on the surface of liquid, presuming non direct contact situation, in which solid particles are separated by water clearings, as it occurs on the surfaces of liquid marbles (see Figs. 3A-B). The values of the effective elastic modulus, which were recently established experimentally in Ref. 35 under deformation of liquid marbles, coated by micrometrically scaled polyethylene particles, are close to 100 Pa, in a good agreement with our estimation.

For nanometric particles, the line tension should be taken into account in Exps. (8) and (9). For example, for hydrophobic colloidal particles of a size $R=10$ nm on water/air interface with above-mentioned parameters we have $\kappa \sim 0.1$, and, putting the average distance between particles equal to their size $R$, the concentration $n \sim 10^{16}$ m$^{-2}$. Then from (9) it follows $E \sim 10$ MPa, that is much more than in the case of micro-scaled particles. This is not surprising; Exp.1 yields for nanoparticles similar values of the elastic modulus. It is also obvious from (9) that in the case of materials with lower values of the contact angle $\theta$, the value of the effective Young modulus will be much lower.

It is noteworthy that the proposed model of transversal elasticity of surfaces coated with colloidal particles implies zero contact angle hysteresis [31-33]. Considering the contact angle hysteresis will call for much more complicated mathematical treatment.

It should be stressed that the proposed mechanism of elasticity will work only at the initial stage of the deformation of droplets coated with colloidal particles (see Figs. 1-2), on "short" time spans of deformation $\tau_1 \cong R/v$, where $v$ is the characteristic velocity of the deformation; at the characteristic times, given by $\tau_2 \cong a/v$, where $a$ is the characteristic dimension of an entire droplet, the mechanisms of elasticity discussed in Ref. 34 will be dominant. The characteristic time of deformation of water droplets bouncing solid substrates was established in Ref. 31 as $\tau_2 \cong \sqrt{\rho a^3 / \gamma}$. This is not surprising, because in the case of bouncing droplets the role of the effective spring stiffness $k$ is played by the surface tension of a bouncing drop. Thus, the effective mechanical scheme of a liquid surface coated with solid colloidal particles looks like as it is depicted in Figure 6 ($\eta$ is the viscosity of the liquid).

## 3. Conclusions

We introduce the mechanisms of elasticity of liquid surfaces coated with colloidal particles in the case when the gravity and capillary interactions are negligible. We treat the situation when solid colloidal particles do not form close packaging and are separated by water clearings. The deformation normal to the surface of a colloidal layer is discussed. When a colloidal particle is displaced from its equilibrium position the pseudo-elastic force following the Hooke-like law arises. This purely interfacial force gives rise to the pseudo-elasticity of the colloidal raft coating a liquid surface.

The effective Young modulus of a surface depends on the interfacial tension, equilibrium contact angle, radius of colloidal particles and their surface density. Eigen-frequency of oscillations of colloidal particles is calculated. For the nanometrically scaled particles the line tension may dominate over interfacial tension effects. The proposed mechanism of elasticity will work only at the initial stage of the deformation, on "short" time spans of deformation.

**Acknowledgements**

Acknowledgement is made to the donors of the American Chemical Society Petroleum Research Fund for support of this research (Grant 52043-UR5).

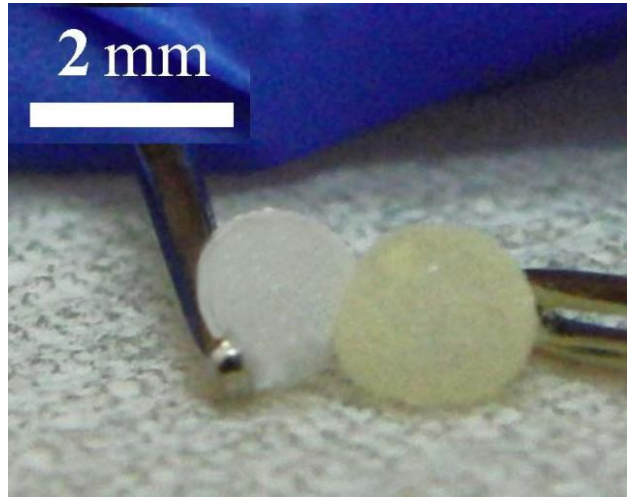

**Fig. 1** 10 µl Teflon (white) and lycopodium (yellow)-coated marbles do not coalesce even when pressed one to another.

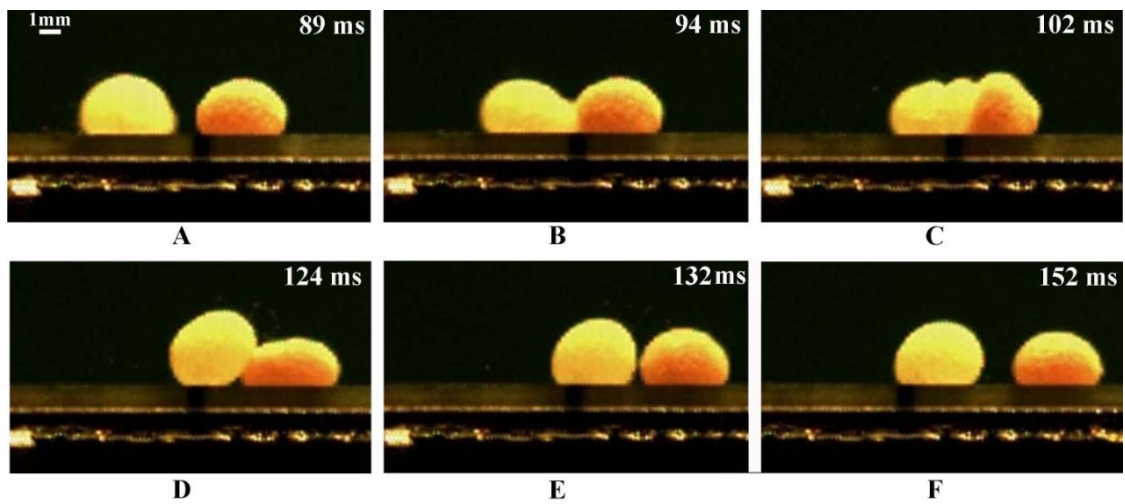

**Fig. 2** Sequence of images illustrating the quasi-elastic collision between lycopodium-coated liquid marbles. Scale bar is 1 mm.

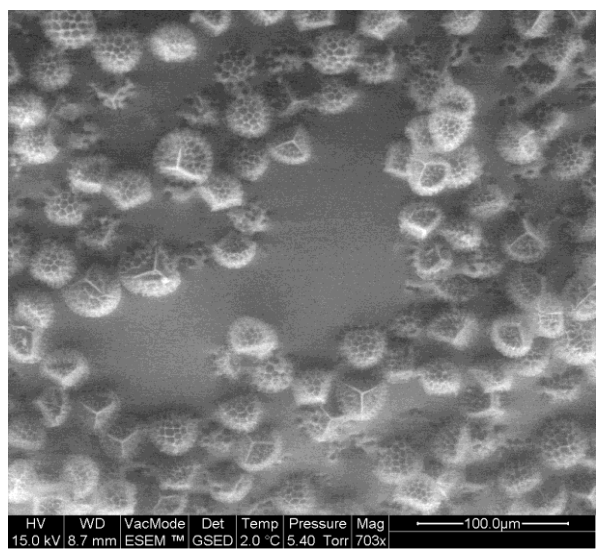

**Fig. 3A** ESEM image of the surface of liquid marble coated with lycopodium particles. Scale bar is 100 μm.

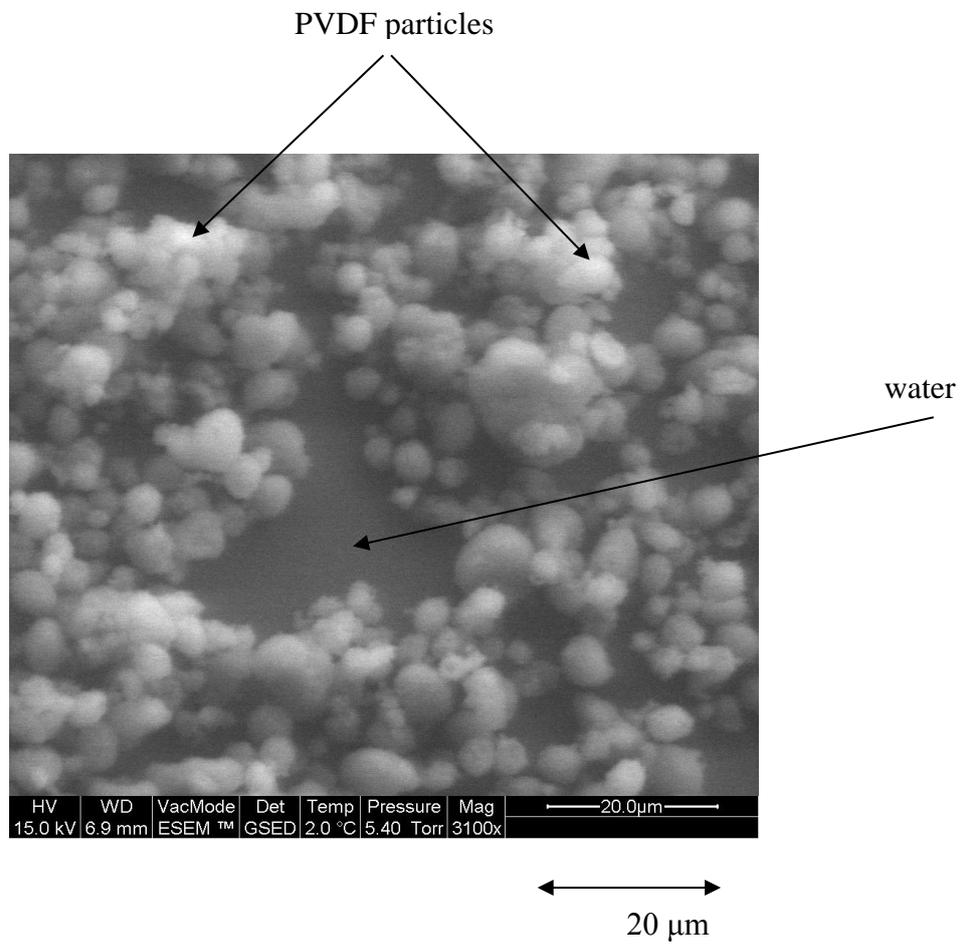

**Fig. 3B** ESEM image of the marble's surface (white spots are aggregates comprising particles of PVDF). Scale bar is 20 μm.

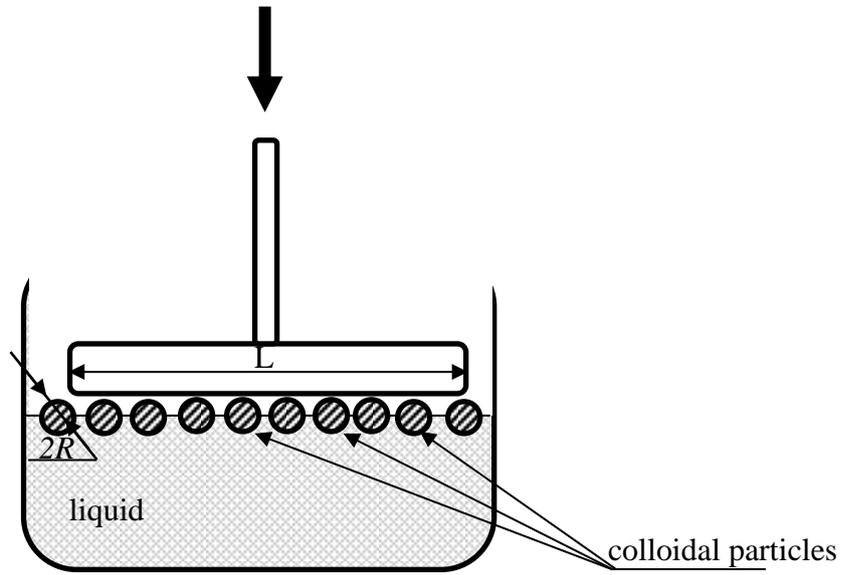

**Fig. 4** The transversal elasticity of a liquid surface, coated by the colloidal layer: the layer is pressed by a plate moving normally to the surface.

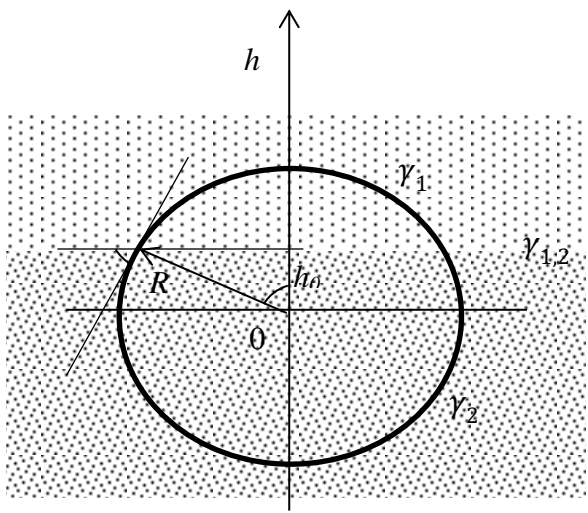

**Fig. 5** Colloidal particle with the radius *R*, located at the interface separating two media.

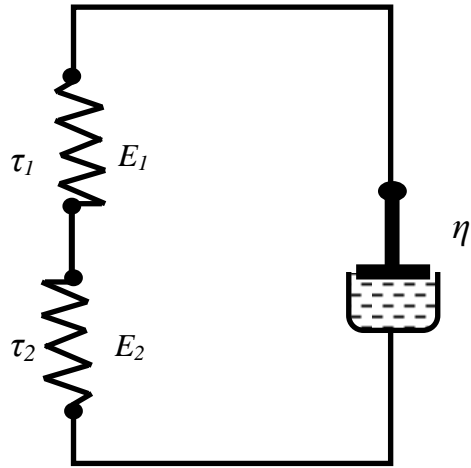

**Fig. 6** Effective mechanical scheme of a liquid surface coated with colloidal particles ($E_1$, $E_2$ are the effective Young moduli, $\eta$ is the viscosity of the liquid).

**TOC Image**

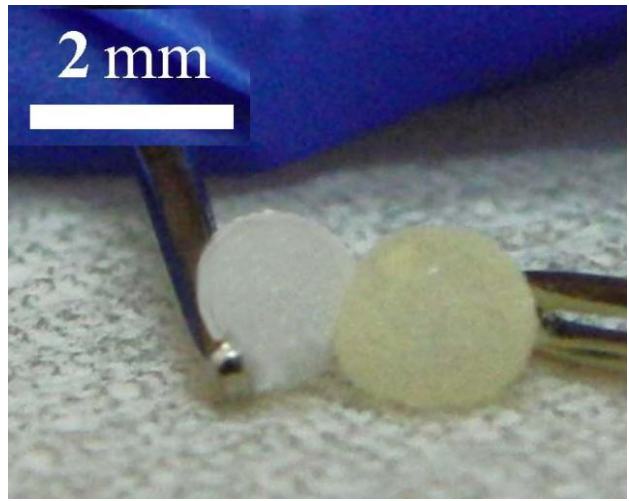

$$E \approx 2\pi n R \gamma_{1,2} |\cos\theta - \kappa\cot^3\theta|$$